\documentclass[aps,pra,twocolumn,groupedaddress,amssymb,amsmath,amsthm]{revtex4-1}
\usepackage{amsfonts}

\usepackage{amssymb}

\usepackage{verbatim}
\usepackage{amsmath}
\usepackage{graphicx}

\newcommand{\bfk}{{\bf k}}
\newcommand{\bfp}{{\bf p}}
\newcommand{\bfq}{{\bf q}}

\newcommand{\ket } [1] {| #1 \rangle}

\newcommand{\expv}[1]{\langle #1 \rangle}

\begin{document}
\title{Variational study of polarons in Bose-Einstein condensates}
\author{Weiran Li and S. Das Sarma}
\affiliation{Condensed Matter Theory Center and Joint Quantum Institute, Department of Physics, University of Maryland, College Park, MD 20742}
\date{\today}
\begin{abstract}
We use a class of variational wave functions to study the properties of an impurity in a Bose-Einstein condensate, i.e., the Bose polaron. The impurity interacts with the condensate through a contact interaction, which can be tuned by a Feshbach resonance. We find a stable attractive polaron branch that evolves continuously across the resonance to a tight-binding diatomic molecule deep in the positive scattering length side. A repulsive polaron branch with finite lifetime is also observed and it becomes unstable as the interaction strength increases. The effective mass of the attractive polaron also changes smoothly across the resonance connecting the two well-understood limits deep on both sides. 
\end{abstract}
\maketitle

\section{Introduction}

Understanding the properties of a single quantum impurity immersed in and interacting with a macroscopic background is an important old topic in many-body physics. Finding out how this bare particle is dressed by the medium when the interaction is turned on reveals the emergent physics of quantum many-body renormalization. One well-known example is an electron moving in a lattice of ions interacting with the phonons of the underlying lattice\cite{MahanSDS,elec1,elec2}. This lattice-dressed electron, universally referred to as a polaron in solid state physics, has properties significantly different from the simple single free band electron (i.e., the bare ``impurity") as it carries a phonon cloud (i.e., is renormalized) in moving through the lattice. While in the weakly interacting limit the polaron simply has a larger renormalized effective mass\cite{MahanSDS} than the free electron (the so-called weak or Fr\"ohlich polaron), the renormalization effect could become extreme in the strongly interacting regime (the so-called strong or Landau-Pekar polaron), where the electron may become self-trapped and localized (i.e., infinite effective mass) completely losing its mobility\cite{elec1,elec2}. Although the polaron problem is an old and extensively studied many-body problem in solid-state physics, recent progress in ultra-cold atomic gases provides an alternative route in studying polaron phenomena in the completely different context of collective atomic physics. The atomic gases have certain advantages in studying the polaron physics as the interaction strength can be tuned easily and precisely by Feshbach resonances\cite{Chin}, thus allowing the continuous tuning of the same system from the weak- to the strong-polaron limit---something impossible to achieve in solid-state materials where the electron-phonon coupling strength is fixed in a given system. Also as one can adjust the shape and the dimension of the atomic clouds as well as choose different types of particles\cite{coldRMP, Rydberg, dipolar}, cold atoms provide a much broader perspective of studying the quantum polaron problem. There have already been experimental studies of Fermi polarons (impurities in fermionic media, and we refer to impurities in Bose media as Bose polarons) in different dimensions in the context of highly polarized Fermi gases\cite{FPLR3, FPLR1, FPLR2}. For Bose polarons, however, it is not until recently that controlling Feshbach resonances in Bose-Fermi mixtures\cite{BFmix1, BFmix2, BFmix3} makes it possible to systematically study a fermionic impurity in a bath of Bose-Einstein condensates (BECs). 

Recent studies of Bose polarons have extensively used the mean-field approach\cite{BPLR1, BPLR2, BPLRP1, BPLR3, BPLR4, BPLR5}, which is strictly accurate only in the weakly interacting regime. These mean-field studies on both homogenous and trapped BECs find polaronic self-trapping in the strongly repulsive regime. It is, however, not clear if the mean-field results on self-trapping are reliable as the interaction effect can be significantly overestimated in the strongly interacting regime. In order to address the properties of the Bose polarons close to the strong-coupling Feshbach resonance appropriately, one must go beyond the mean-field approximation. For the weakly interacting case, a nonperturbative resummation approach has been applied to study the Bose polaron\cite{Demler}. In the more interesting strongly interacting regime, there has been a recent diagrammatic study of the Bose polaron problem over the whole Feshbach resonance\cite{RathSchmidt}, in which the spectral function of the Bose polaron at zero temperature is calculated by non-self-consistent and self-consistent $T$-matrix approximations. The variational theory can also be used to study the polaron problem. It makes no explicit assumption regarding the coupling strength, and is therefore equally applicable to both weak-coupling and strong-coupling situations, but the regime of validity of any variational theory (as well as its quantitative accuracy) is not easy to ascertain. 

In this paper we develop a different variational theory to study the properties of Bose polarons at zero temperature, by generalizing the Chevy ansatz introduced in the Fermi polaron problem\cite{ChevyPLR}. The interaction effect is treated beyond the mean-field level in our theory and thus can be applied to the strongly interacting regime. By writing down a set of trial wave functions and minimizing their energy with respect to the Hamiltonian, we calculate the Bose polaron dispersion relation obtaining the polaronic effective mass at small momentum. We show that the attractive and repulsive branches correspond to two different solutions of the variational energy minimization problem. We also estimate the lifetime of the metastable repulsive polaron directly from the imaginary part of the variational energy minimum, thus obtaining both the effective mass and the lifetime of the Bose polaron on an equal footing. Our results are consistent with those from the diagrammatic approach in Ref. \cite{RathSchmidt} in which the spectral function of Bose polarons at zero temperature is calculated. 

This paper is structured as follows: In Sec. \ref{sec:model} we introduce a very general model for the Bose polaron problem. In Sec. \ref{sec:non} and Sec. \ref{sec:weak} we show our predictions for polarons in the noninteracting and weakly interacting BECs, respectively. Finally in Sec. \ref{sec:con} we summarize our results and discuss the possible extensions within this framework. 

\section{The General Model}  \label{sec:model}
We consider a system of a single impurity (typically a fermion) moving in the background of a Bose-Einstein condensate with finite density $n$. The characteristic length and energy scales are parameterized by the density of the Bose background analogous to the Fermi momentum and energy:  $n\equiv k_F^3/6\pi^2$ and $E_F\equiv \hbar^2k_F^2/2m_b$, where $m_b$ is the mass of the bosons in the BEC. Since we only consider the single-particle behavior of the impurity making its statistics irrelevant, we denote the impurity particle as the ``fermion" for simplicity (although the impurity could also be a boson of different species compared with the BEC---our results will remain the same for a bosonic impurity). The fermion-boson and boson-boson interactions are modeled by contact (zero-range) couplings, and are characterized by the $s$-wave scattering lengths $a_{bf}$ and $a_{bb}$ respectively. Also from now on, we set $\hbar=1$.

\section{Non-interacting BEC} \label{sec:non}
We start with a non-interacting Bose gas where $a_{bb}=0$. The full Hamiltonian takes the form
\begin{eqnarray}\label{eqn:hamiltonian}
&H=\sum_\bfk\epsilon_b(\bfk) b^\dag_\bfk b_\bfk+\sum_\bfk\epsilon_f(\bfk) f^\dag_\bfk f_\bfk+H_{\rm int},\\
&H_{\rm int}=\frac{g_{bf}}{V}\sum_{\bfk,\bfk',\bfq} b^\dag_{\bfk'} f^\dag_{\bfq-\bfk'} f_{\bfq-\bfk} b_\bfk,
\end{eqnarray}
where $\epsilon_{b,f}(\bfk)=k^2/2m_{b,f}$ are the kinetic energy terms for majority bosons and the impurity fermion respectively. The coupling constant $g_{bf}$ in the interaction Hamiltonian is determined by the scattering length $a_{bf}$ between the bosons and the fermion, which can be tuned to any value by a Feshbach resonance. For the zero-range model, the coupling term has an ultra-violet divergence, which is regularized by the following procedure:
\begin{equation}\label{eqn:greg}
\frac{1}{g_{bf}}=\frac{m_{\rm red}}{2\pi a_{bf}}-\frac{1}{V}\sum_\bfk \frac{2m_{\rm red}}{k^2},
\end{equation}
where $m_{\rm red}^{-1}=m_f^{-1}+m_b^{-1}$ is the inverse reduced mass and $V$ is the total volume of the system. The mean-field approximation consists of taking the coupling constant as $g_{bf}^{MF}=2\pi a_{bf}/m_{\rm red}$, which can be either attractive or repulsive depending on the sign of the scattering length.

For a condensate in the thermodynamic limit, we work in the grand-canonical ensemble. The ground state of this non-interacting BEC is a coherent state denoted as $\ket{{\rm BEC_0}}$, with all particles occupying the $k=0$ single-particle state. When the fermion interacts with the bosons, it creates excitations that the $k=0$ bosons have certain probability of being scattered into the $k\neq 0$ states. The fermion itself also gets dressed by these bosonic excitations and becomes a quasi-particle, i.e., the Bose polaron. 

We now write down a class of trial wave functions to include the interplay between the fermion and the bosons:
\begin{equation}\label{eqn:PLRwf}
\ket{\Psi^{(q)}}\sim \left( \psi^{(q)}_0 f_q^\dag+\sum_{k\neq 0}\psi^{(q)}_k f_{q-k}^\dag b_k^\dag  \right)\ket{{\rm BEC_0}}.
\end{equation}
This wave function has a total momentum $q$ and describes a variational state with one single boson excited out of the $k=0$ condensate. 

The theoretical procedure to minimize the energy of this type of variational wave function was first introduced by Chevy in the context of Fermi polarons\cite{ChevyPLR}. It can be generalized for Bose polarons as we show here. To find the local minimum energy for the wave function (\ref{eqn:PLRwf}) with respect to the full hamiltonian (\ref{eqn:hamiltonian}), we use the following differential equation:
\begin{equation}\label{eqn:partial}
\frac{\partial \expv{\Psi|H-E|\Psi}}{\partial \psi^\ast}=0,
\end{equation}
where, by the usual variational principle, the Lagrange multiplier $E$ gives the local energy minimum. In Eq. (\ref{eqn:partial}), $\ket{\Psi}$ is the trial wave function defined by Eq. (\ref{eqn:PLRwf}), and the partial derivatives are with respect to all components of $\psi$ ($\psi_0^{(q)}$ and $\psi_k^{(q)}$) in Eq. (\ref{eqn:PLRwf}). At each total momentum $q$, we have a minimized energy $E(q)$, which provides the dispersion relation of the Bose polaron. The polaronic interaction effect is most naturally extracted from the long-wavelength quasi-particle dispersion behavior. Hence we write down a general expression for the dispersion relation at small momentum $q\sim 0$ as
\begin{equation}\label{eqn:expansion}
E(q)=\Sigma(0)+\frac{q^2}{2m_{\rm eff}}+O(q^4).
\end{equation}
The $\Sigma(0)$ term is the ``self-energy", sometimes referred to as the polaronic binding energy, which gives an overall energy shift from the free particles\cite{selfE}. The prefactor of the quadratic term in $q$ determines the polaronic effective mass of the single impurity fermion dressed by bosonic bath of the BEC $m_{\rm eff}$. 

First we analyze the overall energy shift $\Sigma(0)$ at $q=0$. In general the variational solution contains a real part that represents the interaction strength between the fermion and the condensate bosons, and an imaginary part related to the polaronic lifetime. For the contact interaction Hamiltonian (\ref{eqn:hamiltonian}), the lowest-energy state is always attractive and the attraction increases monotonically as the scattering length crosses the Feshbach resonance from the negative to the positive side. In the limit $k_F a_{bf}\rightarrow 0^+$, the fermion is tightly bound to a single boson forming a diatomic molecule. There is also a metastable branch where $a_{bf}>0$, with an effective repulsive interaction if the system is prepared free of the bound-state molecules. In our calculations these two branches correspond to two different solutions to Eq. (\ref{eqn:partial}). The attractive branch has ${\rm Re}\Sigma_{\rm att}<0$ and the repulsive branch has a positive energy shift ${\rm Re}\Sigma_{\rm rep}>0$.  

\begin{figure}
  \includegraphics[width=0.38\textwidth]{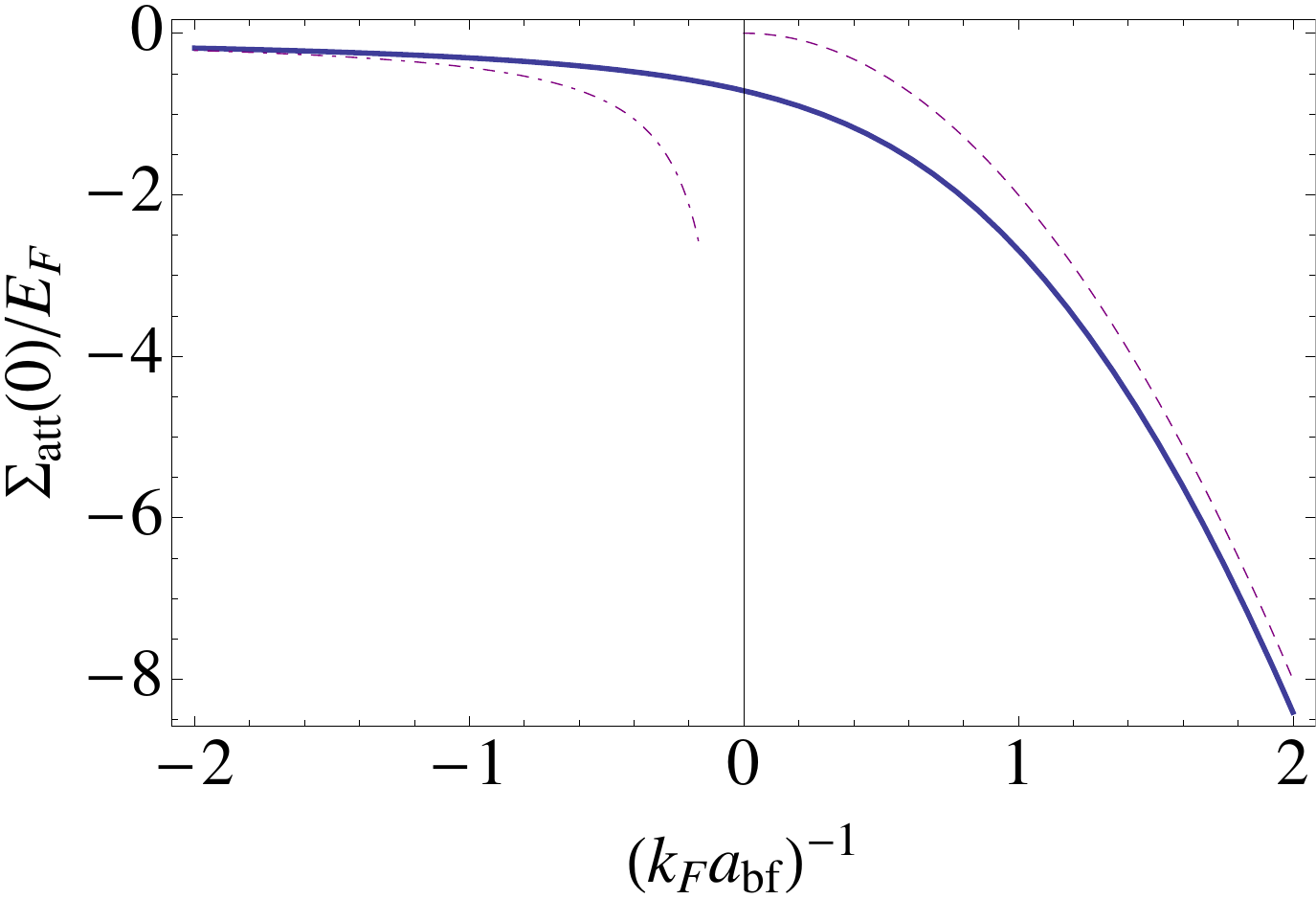}
    \caption{The solid line represents the attractive self-energy $\Sigma_{\rm att}(0)$. The dotted-dashed line in the negative scattering length side represents the Hartree approximation $\Sigma_H=g_{bf}n$ that diverges at the resonance. The dashed curve is the asymptotic bound state energy calculated from the two-body physics $E_b=-1/(m_{\rm red}a_{bf}^2)$, where $a_{bf}>0$.}\label{fig:self-non-att}
\end{figure}

\begin{figure}
  \includegraphics[width=0.38\textwidth]{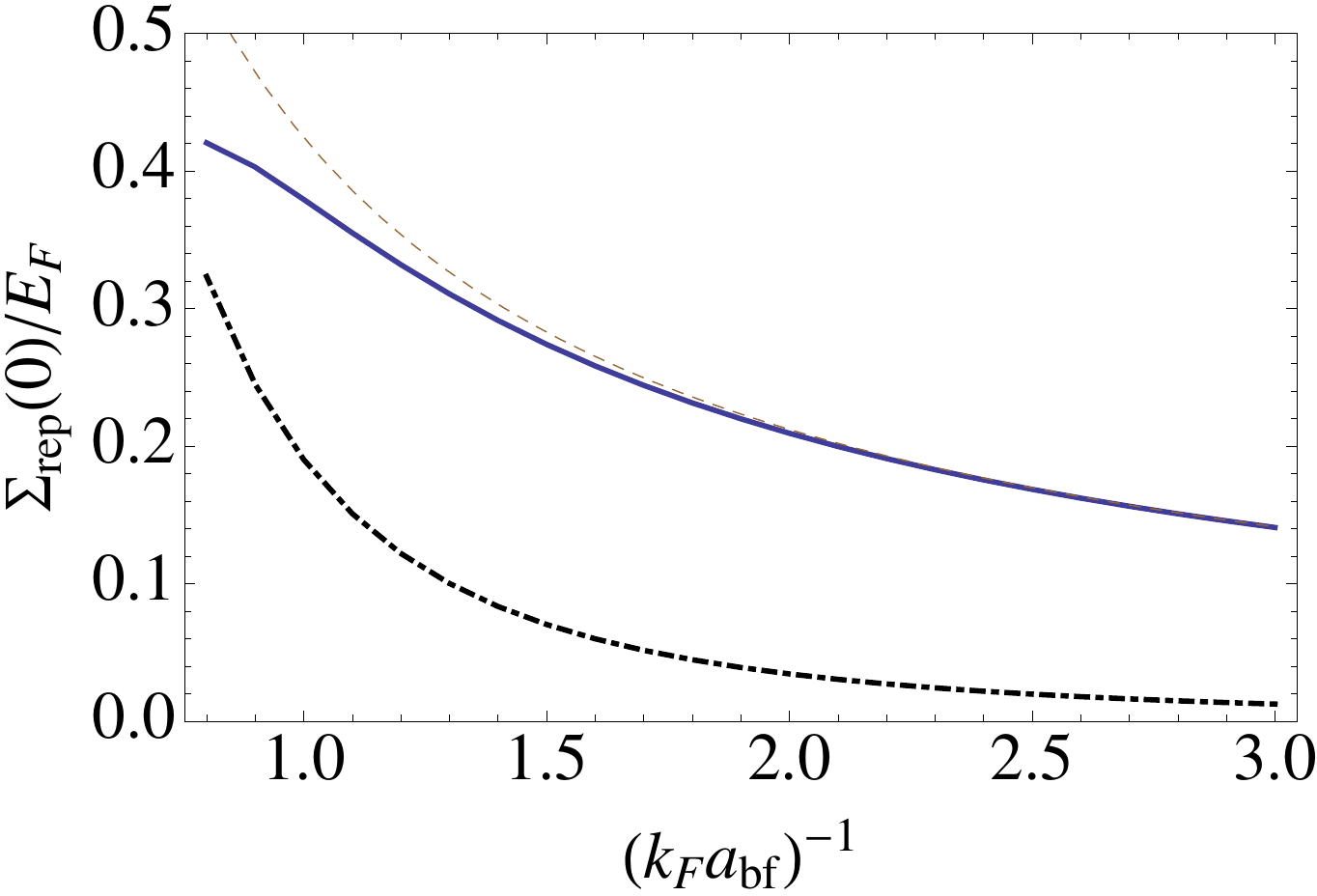}
    \caption{The real part (solid) and the imaginary part (dotted-dashed) of the repulsive polaronic self-energy. The thin dashed line is the asymptotic behavior of the Hartree approximation in the weakly repulsive regime ($a_{bf}\rightarrow 0^+$). In the strongly interacting regime where $k_F a_{bf}\gtrsim  1$, the imaginary part becomes comparable to the real part of the self-energy. The repulsive polaron is no longer a well-defined quasi-particle beyond this point. }\label{fig:self-non-rep}
\end{figure}

The dispersion relations of the attractive and repulsive polarons with equal mass $m_b=m_f$ are shown in Figs. (\ref{fig:self-non-att}) and (\ref{fig:self-non-rep}) respectively. In the attractive branch, the solution of the self-energy $\Sigma_{\rm att}$ is purely real and negative. In the weakly attractive limit, where $(k_F a_{bf})\rightarrow 0^-$, the perturbation theory gives an analytic result $\Sigma_{\rm att}=g_{bf}n=2\pi n a_{bf}/m_{\rm red}$. This is equivalent to the first-order Hartree approximation for the interaction effect. Deep on the other side of the resonance where $(k_F a_{bf})\rightarrow 0^+$, the attraction between the boson and the fermion becomes so strong that the pair of particles form a deep bound molecule. The bound-state energy of this molecule is $E_b=-1/(m_{\rm red} a_{bf}^2)$. As shown in Fig. (\ref{fig:self-non-att}), the attractive solution to Eq. (\ref{eqn:partial}) agrees with the analytic results in both limits. Over the whole process of the crossover, the self-energy changes continuously and smoothly between the two well-understood asymptotic limits. 

For the metastable repulsive branch, the solution of the self-energy contains a positive real part ${\rm Re}\Sigma_{\rm rep}>0$ that characterizes the strength of the repulsion, and a finite imaginary part ${\rm Im}\Sigma_{\rm rep}\neq 0$ that indicates the instability (i.e., the finite lifetime) of this repulsive Bose polaron. The real and imaginary parts are shown in Fig. (\ref{fig:self-non-rep}) on the positive scattering length side of the Feshbach resonance. In the weakly repulsive case where $(k_F a_{bf})\rightarrow 0^+$, the real part of the self-energy of the repulsive branch is very close to the Hartree approximation, namely $\Sigma_{\rm rep}(0)\approx\Sigma_{\rm rep}^{H}=2\pi n a_{bf}/m_{\rm red}>0$. As the scattering length increases, the interaction energy becomes larger deviating from the Hartree result. On the other hand, the imaginary part determines the damping rate and the polaronic lifetime. For a repulsive polaron at zero momentum, the total energy equals the self-energy, so its propagation can be written as $e^{i\Sigma t}=e^{i{\rm Re}\Sigma \cdot t}e^{-{\rm Im}\Sigma\cdot t}=e^{iEt}e^{-t/\tau}$. The polaron is thus well defined only when it has the real part of self-energy much larger than the imaginary part. 

As shown in Fig. (\ref{fig:self-non-rep}), in the weakly repulsive limit far off resonance we have ${\rm Im}\Sigma\ll {\rm Re}\Sigma$, which corresponds to a very stable repulsive polaron. When the scattering length becomes larger close to the resonance, ${\rm Im}\Sigma$ becomes comparable to ${\rm Re}\Sigma$, and the decay process dominates so that there is no well-defined repulsive branch of the Bose polaron. This generic unstable feature arises from the fact that the effective repulsion between atoms only occurs in the metastable state. The ``true ground state" of the atomic cloud is always attractive, with the underlying bound-state molecules populated when the scattering length is positive\cite{instabF1,instabF2,instabB}. 

\begin{figure}
  \begin{center}
  \includegraphics[width=0.38\textwidth]{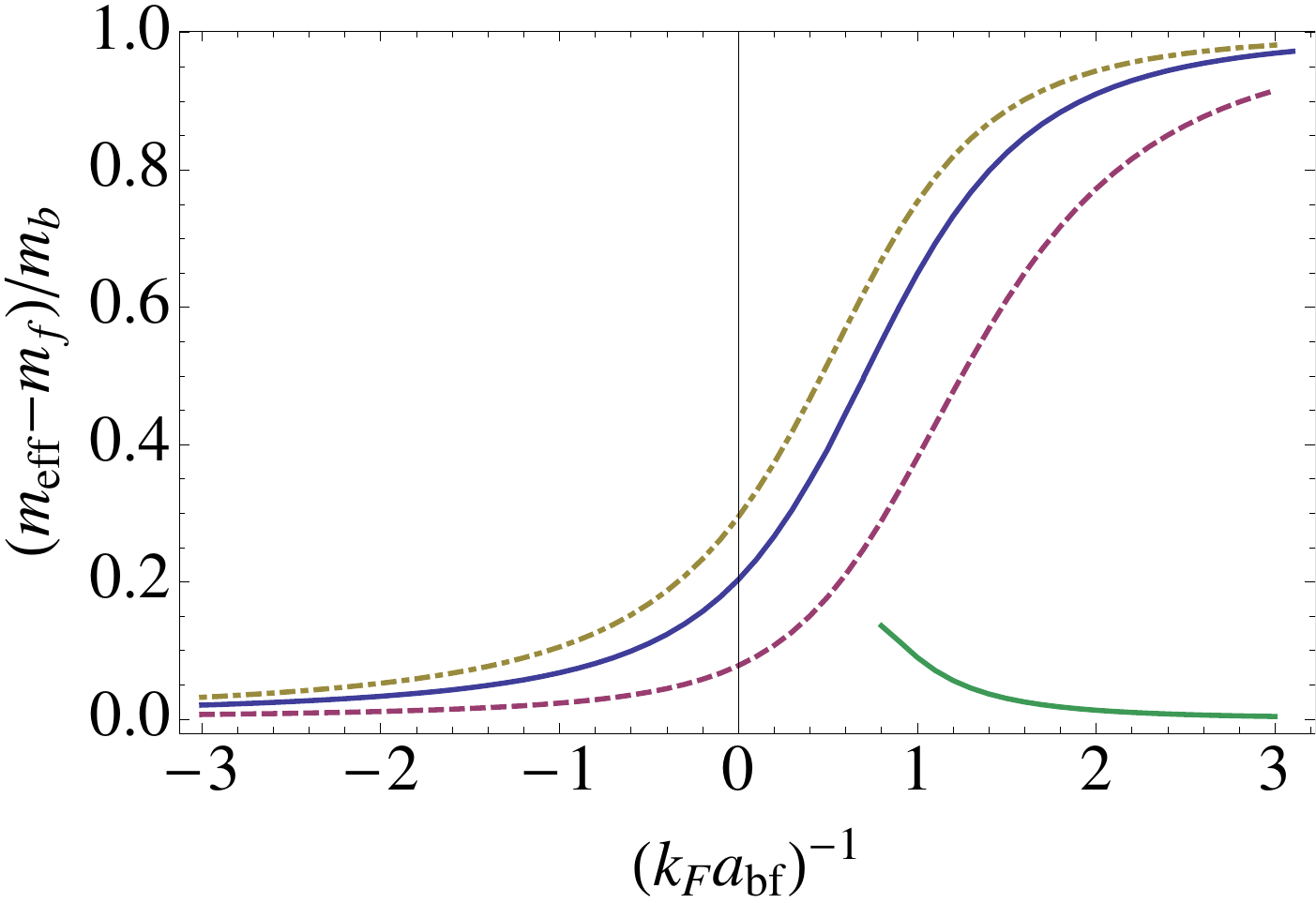}
    \caption{The rescaled dimensionless effective mass renormalization for different mass ratios $\alpha=m_f/m_b$. Three continuous curves between $(k_Fa_{bf})^{-1}=-3$ and $(k_Fa_{bf})^{-1}=3$ are of the attractive polarons. The solid, dashed purple, and dotted dashed brown curves represent $\alpha=1$, $\alpha=0.2$, and $\alpha=5$, respectively. A polaron consists of a heavier impurity has a larger rescaled effective mass renormalization. The green curve that appears in the positive side only represents the effective mass of the $\alpha=1$ repulsive polaron. The repulsive polaron becomes unstable beyond the ending point.}\label{fig:meff-non}
 \end{center}
\end{figure}

The interaction effect also determines the effective mass renormalization of the Bose polaron, which is obtained from the low momentum expansion in Eq. (\ref{eqn:expansion}). We use a rescaled dimensionless value of the effective mass renormalization $(m_{\rm eff}-m_f)/m_b$ to describe the number of bosons contributing to the ``dressing" of the impurity fermion. Its value is shown in Fig. (\ref{fig:meff-non}) for both attractive and repulsive polarons. For the attractive branch, very few bosons contribute to the polarons in the weakly interacting regime. This number gradually increases as the attraction ramps up and finally saturates at $(m_{\rm eff}-m_f)/m_b=1$ deep in the positive scattering length side. It signifies that effectively one single boson contributes to the formation of the polaron in this limit, consistent with the nature of the quasi-particle being a diatomic molecule with an effective mass $m_b+m_f$. We also plot the rescaled effective mass for the situation with different mass ratios between the bosons and the fermion. We find the intuitive result that the heavier impurity attracts more bosons, which can be understood from a Born-Oppenheimer-type argument that lighter particles usually move faster and contribute more to the interaction strength. The repulsive polarons exhibit a similar feature: the effective mass increases as the scattering length comes close to the Feshbach resonance until the quasiparticle no longer remains well defined beyond a certain interaction strength, with the whole process being a smooth crossover with nothing drastic happening anywhere.

All physical properties of the attractive polarons discussed in this section change continuously, and are smoothly connected between weakly interacting and tight molecular limits deep on the two sides of the Feshbach resonance. This is absolutely different from the corresponding situation for the Fermi polaron where a polaron-molecule transition appears\cite{PMtransition, PMtransition2, PMtransition3}. The reason for this difference is that the bosonic ground state we use remains itself after annihilating one macroscopically occupied $k=0$ particle, i.e., $b_0\ket{{\rm BEC_0}}=\sqrt{N_0}\ket{{\rm BEC_0}}$, thus a single trial wave function given by Eq. (\ref{eqn:PLRwf}) is able to describe both the weak polaron and the tight-bound molecule in contrast to the Fermi polaron. 

\begin{figure}
  \begin{center}
  \includegraphics[width=0.38\textwidth]{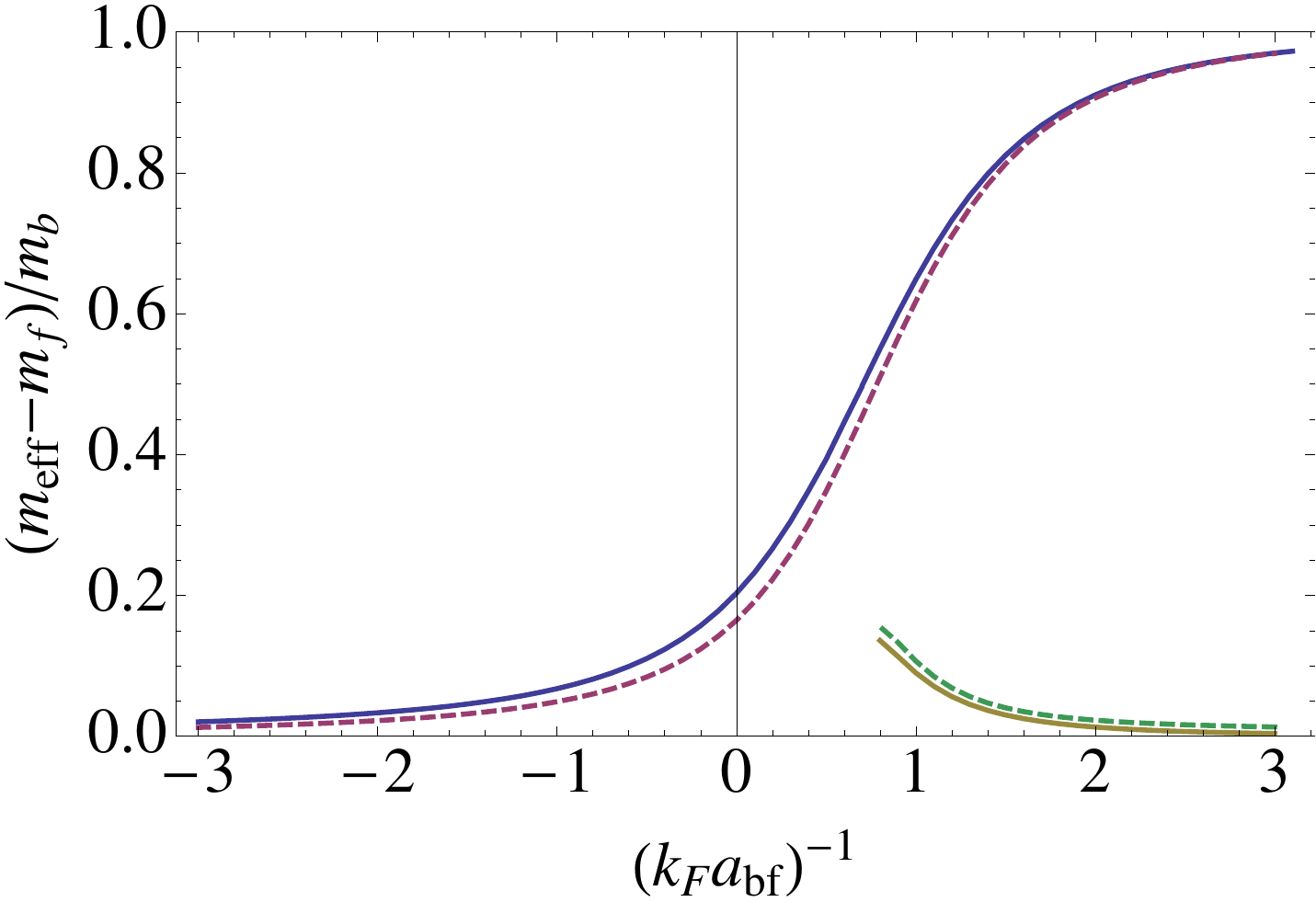}
    \caption{The effective mass of the attractive and repulsive polarons in non-interacting (solid) and weakly interacting (dashed, $k_F a_{bb}=0.2$) BEC backgrounds. The continuous curves across the resonance are the attractive polarons. The repulsive polarons only exist in the positive scattering length side of the resonance.}\label{fig:meff-weak}
 \end{center}
\end{figure}

\section{Weakly interacting BEC} \label{sec:weak}
Now we turn on a weak repulsion between the bosons. The ground state of this weakly repulsive BEC is solved by the mean-field Bogoliubov theory. The depletion of the condensate is given by $n_{ex}/n=8(na_{bb}^3)^{1/2}/3\sqrt{\pi}\approx 0.2(k_F a_{bb})^{3/2}$\cite{PS}, and is negligible when $k_F a_{bb}$ is small. In this limit the elementary excitations in the condensate are the collective Bogoliubov modes which have a dispersion relation given by $\gamma(k)=\sqrt{\epsilon_b(k)\left(\epsilon_b(k)+2g_{bb}n\right)}$. In the Bogoliubov theory the mean-field repulsion has the coupling $g_{bb}=4\pi a_{bb}/m_b$. We generalize the trial wave function of the polarons from the non-interacting case of Eq. (\ref{eqn:PLRwf}) to the following form
\begin{equation}\label{eqn:weaktrial}
\ket{\Phi^{(q)}}\sim \left( \phi^{(q)}_0 f_q^\dag+\sum_{k\neq 0}\phi^{(q)}_k f_{q-k}^\dag \alpha_k^\dag  \right)\ket{{\rm BEC}},
\end{equation} 
where the trial wave function deviates from the original expression (\ref{eqn:PLRwf}) by replacing the noninteracting $\ket{{\rm BEC}_0}$ by the interacting $\ket{{\rm BEC}}$, as well as by replacing the excitations from the $k\neq 0$ free bosons by the corresponding Bogoliubov modes\cite{PS8}.  In the basis of the Bogoliubov modes, the Hamiltonian projected to the trial wave function (\ref{eqn:weaktrial}) is
\begin{eqnarray}
&H=E_g+\sum_\bfk\gamma(\bfk) \alpha^\dag_\bfk \alpha_\bfk+\sum_\bfk\epsilon_f(\bfk) f^\dag_\bfk f_\bfk+H_{\rm int},\\
&H_{\rm int}=\frac{g_{bf}}{V}\Big(N_0\sum_\bfk  f_\bfk^\dag f_\bfk+ \sqrt{N_0}\sum_{\bfk,\bfp}f^\dag_{\bfk+\bfp}f_\bfk \mathcal{R}(p)(\alpha_p+\alpha^\dag_{-p}) \nonumber\\ &+\sum_{\bfk,\bfk',\bfq} f^\dag_{\bfp+\bfk'-\bfk} f_{\bfp}\mathcal{D}(k,k')\alpha^\dag_{\bfk'}  \alpha_\bfk \Big),
\end{eqnarray}
where $\mathcal{R}(p)=u_p-v_p$, $\mathcal{D}(k,k')=u_ku_{k'}+v_kv_{k'}$, and $u_k, v_k$ are the coefficients of the Bogoliubov transformation. These factors arise as the contact interaction hamiltonian undergoes a Bogoliubov rotation. The interaction terms which contain $\alpha\alpha$ or $\alpha^\dag\alpha^\dag$ are absent here since our trial wave function only contains one single Bogoliubov excitation. $E_g$ is the ground-state energy of the weakly interacting BEC and is neglected in the following steps since it trivially shifts the energy of the system.  

Following the same procedure used in the previous section to minimize the energy, the self-consistent equation to solve the polaron energy spectrum in the weakly interacting BEC reads
\begin{equation}\label{eqn:consis}
E(q)-\epsilon_f(q)=n\left(  \frac{1}{g_{bf}}-\frac{1}{V}\sum_\bfk \frac{\mathcal{R}(k)^2 \mathcal{I}(k)}{E-\gamma(k)-\epsilon_f(\bfq-\bfk)}   \right)^{-1}.
\end{equation}
It differs from the energy equation in Ref. \cite{ChevyPLR} by replacing the free boson dispersion $\epsilon_b(k)$ by the corresponding Bogoliubov dispersion $\gamma(k)$, as well as introducing additional factors $\mathcal{R}(k)$ and $\mathcal{I}(k)$. We now evaluate $\mathcal{I}(k)$ from its explicit form:
\begin{equation}
\mathcal{I}(k)=\frac{\psi_0+\sum_{\bfk'}\mathcal{D}(k,k')\psi(k')/\mathcal{R}(k) }{\psi_0+\sum_{\bfk'}\mathcal{R}(k')\psi(k')}.
\end{equation}
As both the numerator and the denominator are ultraviolet divergent ($\psi_k\sim k^{-2}$ at large momentum), the function $\mathcal{I}(k)$ is universally determined by the high momentum tails of $\mathcal{D}$ and $\mathcal{R}$:
\begin{equation}
\mathcal{I}(k)={\rm lim}_{k'\rightarrow \infty}\frac{\mathcal{D}(k,k')/\mathcal{R}(k)}{\mathcal{R}(k')}=\frac{u_k}{\mathcal{R}(k)}.
\end{equation}
By substituting this explicit expression of $\mathcal{I}(k)$ to Eq. (\ref{eqn:consis}) one gets the final self-consistent equation of the polaronic energy. 

The repulsion between the bosons modifies the physical quantities of the Bose polarons, including the self-energy and the effective mass. Here we consider the effective mass for both attractive and repulsive branches as an example to see this effect from boson-boson interactions. We choose the interaction between the bosons to be $k_F a_{bb}=0.2$, and we assume that the bosons have the same mass as the fermion. As shown in Fig. (\ref{fig:meff-weak}), in the attractive branch the effective mass is smaller in the interacting BEC than in the noninteracting BEC. For the repulsive polarons, by contrast, the boson-boson interaction increases the effective mass. A simple physical picture to explain this is that although the boson-fermion attraction tends to bring bosons to closer proximity of the fermion, the repulsion between bosons counteracts this effective binding process. In contrast, for the repulsive branch the boson-boson repulsion obviously plays an opposite role and always enhances the number of bosons in the dressing. Basically, the presence of an inter-boson repulsive interaction in the BEC either suppresses or enhances the polaronic fermion-boson interaction effect depending on whether the polaron is in the attractive or the repulsive branch respectively.

\section{Conclusion and Outlook}\label{sec:con}
In this paper we study the attractive and repulsive Bose polarons in Bose-Einstein condensates using a variational approach. The trial wave functions we propose contain a single excitation of bosons out of the condensate background. The interaction energy, lifetime, as well as the effective mass of the Bose polarons can be all captured by finding the energy of the variational wave function. We do not find any polaronic self-trapping localization (i.e., a divergence of the quasi-particle effective mass) transition at a finite value of the coupling strength in contrast to the mean-field results. Possible modifications in these properties due to higher-order excitations remain an interesting question for the future. Reference \cite{RathSchmidt} finds a 25\% change in the polaron energy from multiple excitations in the BEC, while in the corresponding fermionic case the higher-order excitations are of much less quantitative significance\cite{FPLRcancel}. One conclusion in Ref. \cite{RathSchmidt}, which is consistent with our variational theory, is that the weak-polaron continuously goes over to the tightly-bound molecule (i.e., the strong-polaron in this context) as the coupling increases without any obvious self-trapping transition.  This is a robust conclusion both in our variational theory and in the diagrammatic theory \cite{RathSchmidt}.

Another interesting topic for future study is the possible mapping between the variational and the diagrammatic approaches. For Fermi polarons, it was shown that the Chevy-type variational approach with one single excitation is precisely equivalent to the $T$-matrix approximation in the diagrammatic theory\cite{equiv}. A similar exact correspondence between the two approaches may also exist in the Bose polaron case, but it is yet to be proven theoretically.

\section{Acknowledgments}
WL gratefully acknowledges discussions with Johannes Hofmann, Xiaopeng Li and Jesper Levinsen. This work is supported by NSF-JQI-PFC and ARO-Atomtronics-MURI.


\begin{thebibliography}{9}
\bibitem{selfE}
Here the definition of ``self-energy" is equivalent to the self-energy approximation in Green's function method. The pole of the Green's function in the frequency domain corresponds to the energy solution in this variational approach.
\bibitem{PS8}
Here we use the number-conserved representation of Bogoliubov modes introduced in Ch. 8 of Ref. \cite{PS}, which only create collective excitations (phonon modes) but conserves the particle number. The wave function (\ref{eqn:weaktrial}), in this situation, describes a state that does not have any number fluctuation. Consequently the variational equation does not need to include the chemical potential term.
\end{thebibliography}
\end{document}